\documentclass[twocolumn, times]{aastex62}
\usepackage{amsmath}

\begin{document}

\title{Mode changing and giant pulses in the millisecond pulsar PSR B1957+20}

\author[0000-0002-6317-3190]{Nikhil Mahajan}
\affil{Department of Astronomy and Astrophysics, University of Toronto, 50 St. George Street, Toronto, ON M5S 3H4, Canada}

\author[0000-0002-5830-8505]{Marten H. van Kerkwijk}
\affil{Department of Astronomy and Astrophysics, University of Toronto, 50 St. George Street, Toronto, ON M5S 3H4, Canada}

\author{Robert Main}
\affil{Department of Astronomy and Astrophysics, University of Toronto, 50 St. George Street, Toronto, ON M5S 3H4, Canada}
\affil{Canadian Institute for Theoretical Astrophysics, University of Toronto, 60 St. George Street, Toronto, ON M5S 3H8, Canada}
\affil{Dunlap Institute for Astronomy and Astrophysics, University of Toronto, 50 St George Street, Toronto, ON M5S 3H4, Canada}

\author{Ue-Li Pen}
\affil{Canadian Institute for Theoretical Astrophysics, University of Toronto, 60 St. George Street, Toronto, ON M5S 3H8, Canada}
\affil{Dunlap Institute for Astronomy and Astrophysics, University of Toronto, 50 St George Street, Toronto, ON M5S 3H4, Canada}
\affil{Perimeter Institute for Theoretical Physics, 31 Caroline Street North, Waterloo, ON N2L 2Y5, Canada}
\affil{Canadian Institute for Advanced Research, 180 Dundas Street West, Toronto, ON M5G 1Z8, Canada}

\correspondingauthor{Nikhil Mahajan}
\email{mahajan@astro.utoronto.ca}

\begin{abstract}
Many radio pulsars have stable pulse profiles, but some exhibit mode changing where the profile switches between two or more quasi-stable modes of emission. So far, these effects had only been seen in relatively slow pulsars, but we show here that the pulse profile of PSR B1957+20, a millisecond pulsar, switches between two modes, with a typical time between mode changes of only 1.7 s (or $\sim\!1000$ rotations), the shortest observed so far. The two modes differ in both intensity and polarization, with relatively large differences in the interpulse and much more modest ones in the main pulse. We find that the changes in the interpulse precede those in the main pulse by $\sim\!25$ ms, placing an empirical constraint on the timescale over which mode changes occurs. We also find that the properties of the giant pulses emitted by PSR B1957+20 are correlated with the mode of the regular emission: their rate and the rotational phase at which they are emitted both depend on mode. Furthermore, the energy distribution of the giant pulses emitted near the main pulse depends on mode as well. We discuss the ramifications for our understanding of the radio emission mechanisms as well as for pulsar timing experiments.
\end{abstract}


\section{Introduction} \label{sec:intro}

Radio pulsars exhibit variability in emission on a wide range of timescales, from extremely short bursts like giant pulses to long-term changes in the emission profile. Most are poorly understood; though for some, correlations with possible physical relevance have been found. For instance, giant pulses seem to occur preferentially in pulsars that have a high magnetic field strength at the light cylinder radius, $B_\mathrm{LC}>10^5\,$G (\citealt{Johnston2004, Bilous2015}; here, the light cylinder radius is $r_\mathrm{LC}\equiv{cP/2\pi}$). Recent searches for giant pulses have had success by targeting pulsars with high $B_\mathrm{LC}$, finding them in the millisecond pulsars \object{PSR J1823-3021A} \citep{Knight2005}, \object{PSR J0218+4232} and \object{PSR B1957+20} \citep{Joshi2004, Knight2006}. 

On the other hand, for ``mode changing'' and ``nulling'' -- where pulsars switch between two or more quasi-stable modes of emissions (with one of the states being an off state in the case of ``nulling'') -- no obvious correlations with pulsar properties have been found (e.g., \citealt{Wang2007}). Mode changing and nulling seem intimately related \citep{Leeuwen2002} and are thought to be manifestations of the same phenomenon: a global reconfiguration of the current flow in the pulsar's magnetosphere \citep{Kramer2006, Timokhin2010}. They have also been linked to drifting subpulses (e.g., \citealt{Redman2005, Rankin1986}), although with less clear a physical picture.

While pulse-to-pulse intensity modulations and drifting subpulses have been observed in millisecond pulsars \citep{Edwards2003}, mode changing and nulling have, thus far, only been observed in normal (or slow) pulsars. This could reflect selection biases, however, since few extensive single-pulse studies of millisecond pulsars have been done \citep{Jenet2001, Bilous2012, Liu2015, Liu2016}.

Here, we show that the millisecond pulsar \object{PSR B1957+20} shows mode changing, and that the properties of the giant pulses we found earlier (in the same data; \citealt{Main2017}) show correlations with the modes.

\section{Observations} \label{sec:observations}

We observed PSR B1957+20, a $1.6$ ms pulsar in an eclipsing binary, for over $9$ hours in four daily $\sim\!2.4$ hour sessions on 2014 June 13 -- 16 at the Arecibo Observatory (as part of European VLBI Network program GP 052), using the $327\,\mathrm{MHz}$ receiver. Details of the observations and the baseband data obtained from it can be found in \cite{Main2017}. For our analysis here, we first coherently dedispersed the left- and right-circular polarization baseband data using a dispersion measure of $29.1162{\rm\,pc\,cm^{-3}}$ \citep{Main2017}. Next, we squared them individually and formed the Stokes parameters $I$ and $V$ (PSR B1957+20 shows very little linear polarization; \citealt{Fruchter1990}), averaging over the three $16\,\mathrm{MHz}$ sub-bands in which the pulsar is well detected (frequency range $311.25-359.25\,\mathrm{MHz}$). We convert to flux units using a nominal system temperature of $120\,\mathrm{K}$ and antenna gain of $10\,\mathrm{K/Jy}$ \footnote{\url{http://www.naic.edu/~astro/RXstatus/327/327greg.shtml}}. As the data was relatively clean, we did not filter for radio-frequency interference. We ignore data taken in and around the eclipse due to observed lensing effects \citep{Main2018} that make it difficult to characterize giant pulses and mode changing.

We find that the integrated flux in the pulse profile exhibits slight variability due to interstellar scintillation and antenna gain drifts. To compensate for these, we normalize the profiles using the main pulse flux (which is not affected much by mode changing) smoothed over a timescale of $10$ s -- chosen to be shorter than to the scintillation timescale of $\sim\!84\,$s \citep{Main2017} yet longer than the timescale of $\sim\!1.7\,$s on which the mode changes occur.

\begin{figure*}
\centering
\includegraphics[width=0.75\textwidth,trim=0 0 0 0,clip]{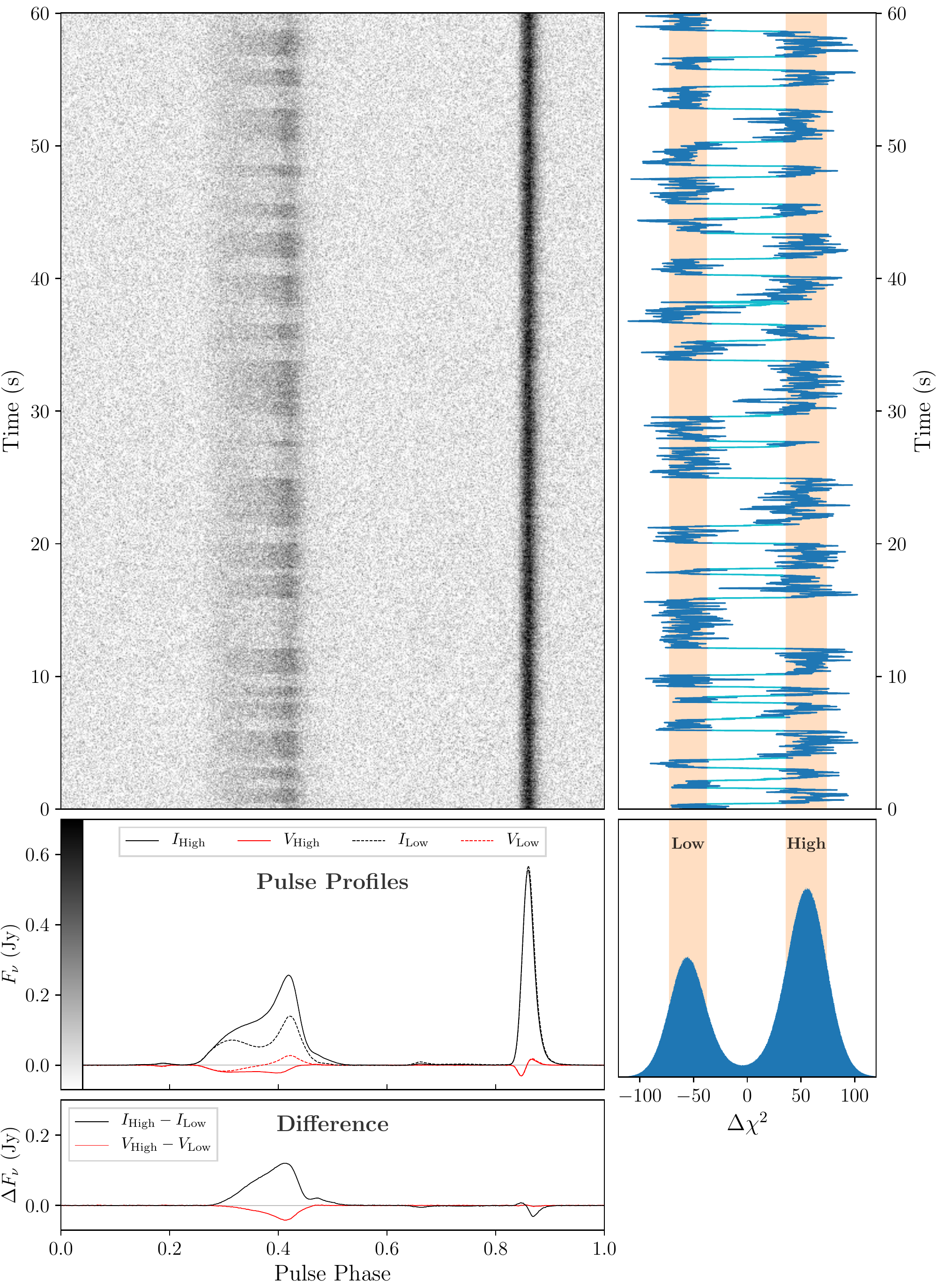}
\caption{{\it (Top Left):} Pulse profiles as a function of time for 1 minute of data, with each row the average of 50 pulse periods. The color-scale corresponds to the vertical axis of {\it Middle Left} panel and a corresponding colorbar is shown on the panel. {\it (Top Right)} The 50-profile $\Delta\chi^2$ for the corresponding profiles on the left, with parts that we define as ``transitions'' shown in cyan. {\it (Bottom Right):} Histogram of the 50-profile $\Delta\chi^2$ for the entire dataset. The orange bars indicate the $\pm 1\sigma$ range around the two peaks (as determined from a fit with the sum of two normal distributions). {\it (Middle Left):} Pulse profiles in the High and Low modes, in Stokes $I$ and $V$. {\it (Bottom Left):} Difference between the High and Low mode profiles, in $I$ and $V$.\label{fig:mode}}
\end{figure*}

\section{Mode Changing}\label{sec:modechanging}

Our data show clear evidence for mode changing in \object{PSR B1957+20} (see Figure \ref{fig:mode}), with the pulse profile switching on timescales of seconds between two quasi-stable modes, which differ in both intensity and circular polarization. Hereafter, we will refer to the two modes as the ``High'' and ``Low'' mode (for the more and less energetic mode, respectively).

The mode changing affects the interpulse most (pulse phase $0.25$--$0.55$ in Figure~\ref{fig:mode}), but small changes are also seen for the main pulse (pulse phase $0.82$--$0.93$) which shifts to a later phase, arriving $937\pm46\,\mathrm{ns}$ later in the Low mode than in the High mode. Furthermore, a weaker pulse component near pulse phase $0.65$ is halved in power in the High mode.

\subsection{Mode Metric}\label{ssec:metric}

In order to determine mode changing properties such as the timescale, we need a quantitative metric for whether an individual pulse belongs to the Low or the High mode. We start by defining $\chi^2$ for each mode,
\begin{equation} \label{eq:1}
\chi^2_{i,\mathrm{mode}}=\sum_{\phi,P}\frac{\left(p_{i,P}(\phi)-p_{\mathrm{mode},P}(\phi)\right)^2}{\sigma_{P}(\phi)^2},
\end{equation}
where $p_{i, P}(\phi)$ is an individual pulse profile for polarization $P$, $p_{\mathrm{mode},P}(\phi)$ the average profile for a given mode in that polarization, and $\sigma_P(\phi)$ the standard deviation from the mean of the entire dataset, calculated for each phase bin $\phi$ separately (this includes effects due to mode changing, though these are small: background noise accounts for 99\% of the variance).

Then, we define a ``mode metric'',
\begin{equation} \label{eq:2}
\Delta\chi^2_{i}=\chi^2_{i,\mathrm{Low}}-\chi^2_{i,\mathrm{High}}.
\end{equation}
For normally distributed data, $\Delta\chi^2$ is twice the logarithm of the likelihood ratio between the high and low modes. Note that, since $\Delta \chi^2$ is just a difference between $\chi^2$ for the same data, it is independent of phase binning (as long as one resolves the profiles well).

For an individual pulse, the expectation values of our metric are $\langle\Delta\chi^2\rangle=\pm\sum((p_{\mathrm{Low}}-p_{\mathrm{High}})/\sigma)^2=\pm1.12$. The expected uncertainty is $\langle\sigma^2_{\Delta\chi^2}\rangle^{1/2}\equiv2|\langle\Delta\chi^2\rangle|^{1/2}=2.12$, which implies that we cannot determine the mode of an individual pulse. Instead, we use a running sum over 50 pulses (chosen to get a sufficiently high signal-to-noise ratio of $\sim\!7$), i.e., we assume implicitly that if an individual pulse is preceded and succeeded by pulses in one mode, then it is itself more likely to be a part of that mode.

An example of the 50-profile metric is shown in Figure~\ref{fig:mode}, as is a histogram for our whole dataset of $\sim\!15$ million pulses. Both show a clearly bi-modal distribution. We fit the histogram with a sum of two normal distributions, finding centers for the low and high mode of $\Delta\chi^2=-55.08\pm0.08$ and $55.03\pm0.05$, respectively, and associated widths of $\sigma_{\Delta\chi^2}=17.72\pm0.08$ and $19.15\pm0.05$. Given the presence of some contamination by mode transitions, the centers agree reasonably with the expected values of $\pm1.12\times50=\pm56.0$. The widths, however, are larger than the expected $2\sqrt{56.0}=15.0$, suggesting additional variability, especially in the high mode.

Even smoothed over 50 pulses, it is not always obvious to which mode an individual 50-pulse average belongs. Indeed, from Figure~\ref{fig:mode}, it seems clear there are times in which the pulsar is transitioning. We quantify such transitions as periods in between a High Mode and Low Mode, or vice-versa, where the smoothed metric goes directly from the inner $1\sigma$ boundary of one mode to the inner $1\sigma$ boundary of the other mode without crossing back over those boundaries in between.\footnote{Our method of identifying transitions implies that, effectively, we ignore the possibility that the pulsar might start to transition from one mode to the other, but then returns.} Once we have determined where all the transitions are (see Figure~\ref{fig:mode}), it becomes trivial to associate all profiles outside of transitions with either the High or the Low mode.

It is important to note that the durations of the transitions do not necessarily represent the true timescales on which mode changes occur. Instead, the transition lengths are set primarily by the length of the smoothing filter applied to the metric. In the context of this paper, it is more useful to think of the transitions as time spans for which the mode is indeterminate.

\subsection{Mode Fractions and Timescale} \label{ssec:timescale}

\begin{figure}
\centering
\includegraphics[width=0.45\textwidth,trim=0 0 0 0,clip]{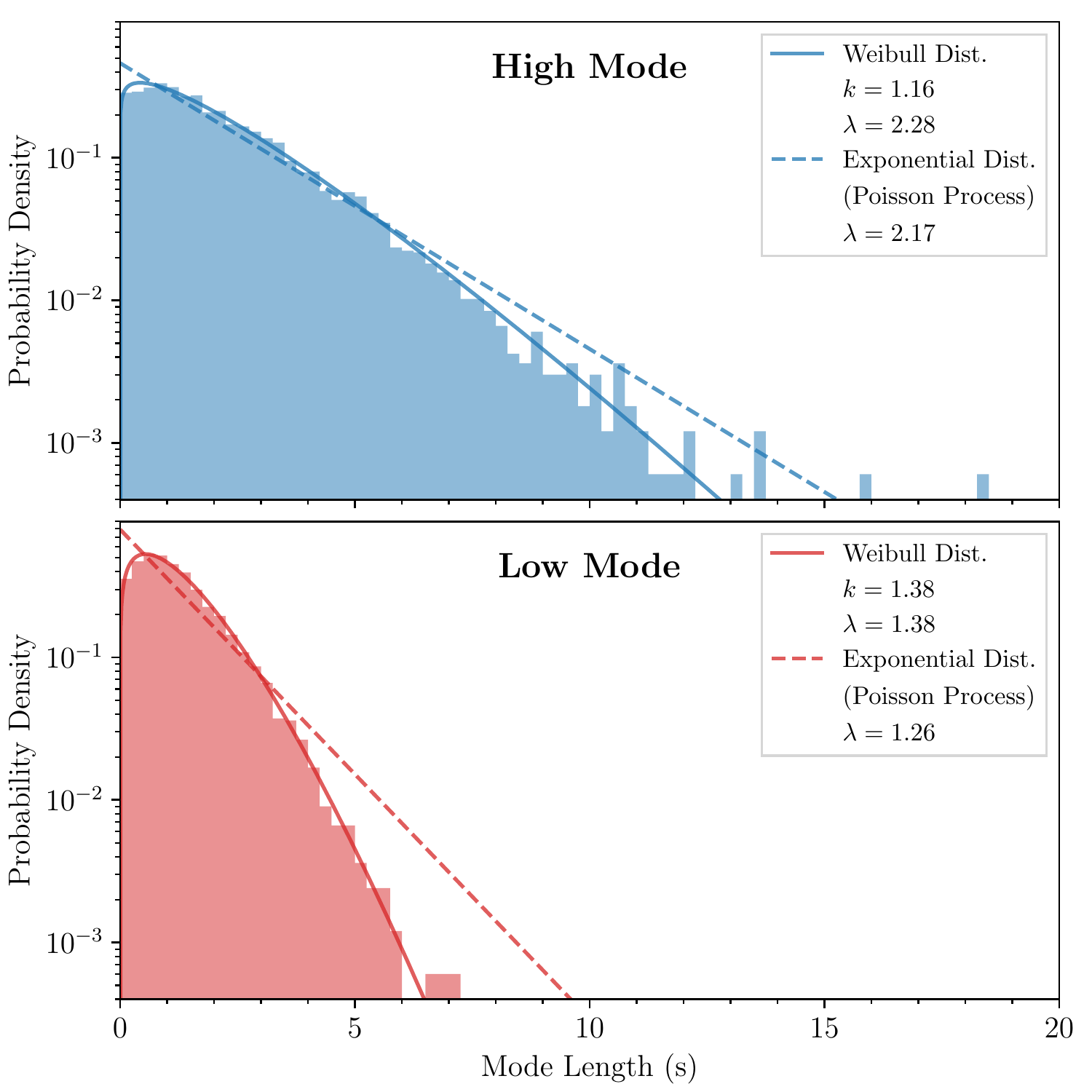}
\caption{Duration distributions for the High {\it (Top)} and Low {\it (Bottom)} modes. The best-fit Weibull and Exponential distributions for each mode are overdrawn, with maximum-likelihood parameters provided in the legend.\label{fig:durations}}
\end{figure}

In our observations, the pulsar is in the High mode $\sim\!60\%$ of the time, in the Low Mode $\sim\!35\%$ of the time; the remaining $\sim\!5\%$ of the time, it is in a transition (this is likely an over-estimate, given our smoothing filter). On average, the time between transitions is $1.7\,$s, or $\sim\!1000$ pulse periods.	

Since the pulsar is in the high mode for a larger fraction of the time, the average time between transitions in the high mode ($2.17\,$s) is proportionally longer than that in the low mode ($1.26\,$s). The duration distributions of the two modes (Figure~\ref{fig:durations}) are well-fit by a Weibull distribution, which has a probability density function,
\begin{equation}
f(x;k,\lambda)=\frac{k}{\lambda}\left(\frac{x}{\lambda}\right)^{k-1}e^{-(x/\lambda)^k}.
\end{equation}
Using maximum-likelihood estimation, we find $k=1.16$ for the High Mode and $k=1.38$ for the Low Mode. A $k > 1$ implies that the probability of a mode change occurring increases with time, the longer the pulsar is in a mode. A simple Poisson process, where the probability of a mode change occurring is time-invariant, produces an exponential distribution ($k = 1$) which does not fit the observed duration distributions well, as shown in Figure~\ref{fig:durations}.

\subsection{Correlations between Main and Inter-pulse} \label{ssec:mpip}

\begin{figure}
\centering
\includegraphics[width=0.45\textwidth,trim=0 0 0 0,clip]{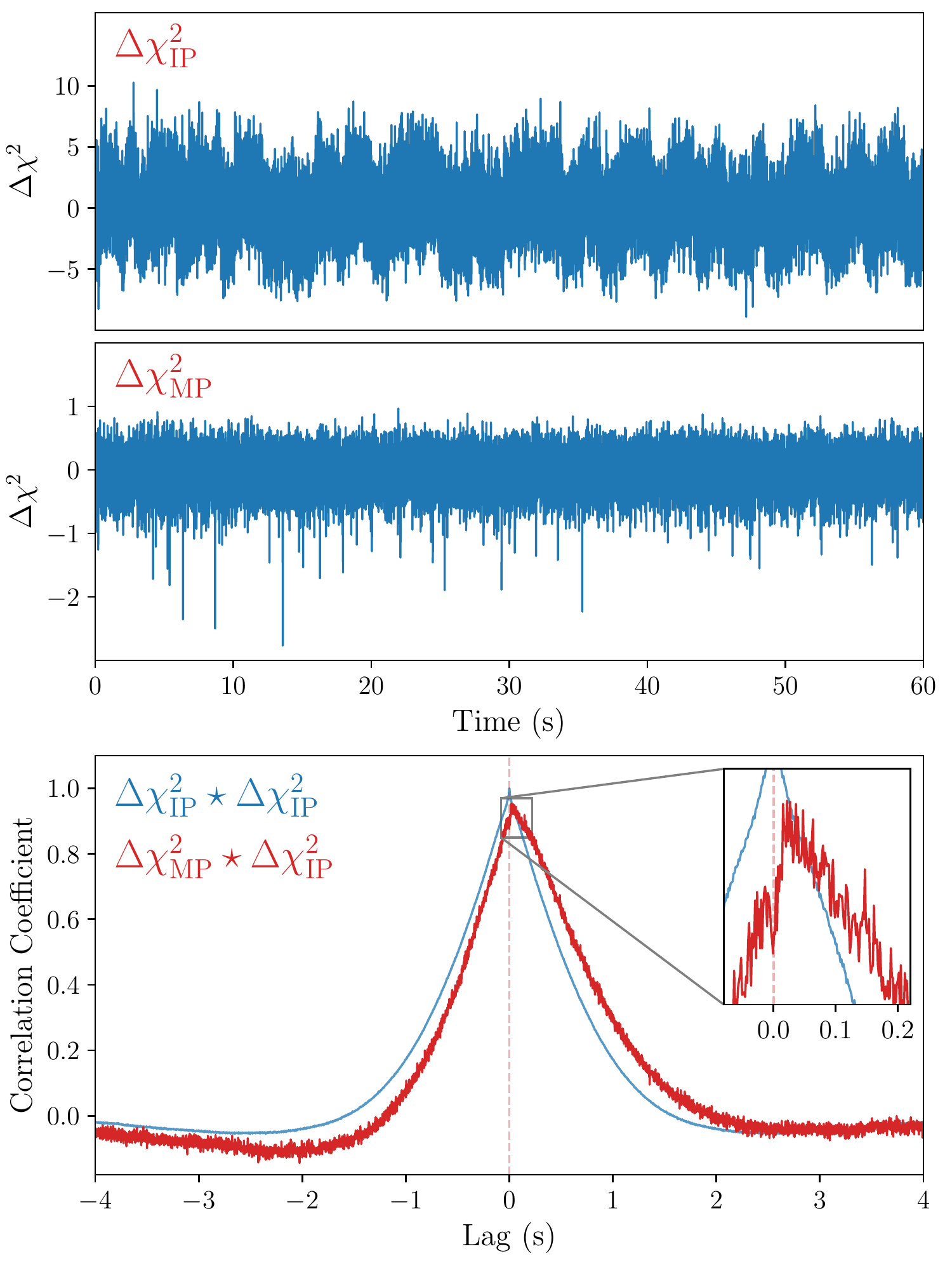}
\caption{Single-pulse mode metrics for the interpulse {\it (Top)} and the main pulse {\it (Middle)} for the 1-minute chunk of data shown in Figure \ref{fig:mode}. The spikes on the main-pulse metric are due to giant pulses. {\it (Bottom)}: Auto-correlation of $\Delta\chi^2_\mathrm{IP}$ and cross-correlation between $\Delta\chi^2_\mathrm{MP}$ and $\Delta\chi^2_\mathrm{IP}$ across the entire dataset, corrected for the contribution of measurement noise to the two metrics. One sees that the main pulse mode metric is delayed relative to that of the interpulse.\label{fig:corr}}
\end{figure}

Since the interpulse and main pulse respond to mode changing in different ways, we generate separate mode metrics for them, as defined in Eqs.~\ref{eq:1} and \ref{eq:2}, with the sum taken over the relevant phase region. A 60\,s snapshot of these two metrics, for individual pulses, is shown in the top two panels of Figure \ref{fig:corr}. One sees that even for the individual pulses, the interpulse metric gives clear evidence for mode changing, but for the main pulse, which only changes subtly between modes, the changes cannot be seen directly.

Nevertheless, from a cross-correlation between the two metrics across our full dataset, it is clear that the metric for the main mode also contains the mode changes: as can be seen in the bottom panel of Figure \ref{fig:corr}, the two are correlated, with a correlation coefficient near unity. In detail, the peak of the cross-correlation is offset from zero lag, and we infer that the main pulse metric lags the interpulse metric by $\sim\!25\,\mathrm{ms}$, or $\sim\!15$ pulsar rotations. This suggests that the physical driver of the mode changes is near(er) the region responsible for the interpulse, and that the timescale for the mode changing effects to propagate across the pulsar's magnetosphere is of order tens of ms.

We also note that the cross-correlation is asymmetric, being skewed towards positive delay. We believe this is real, but have no clue what its physical interpretation might be.

\section{Giant Pulses Revisited} \label{sec:gps}

\begin{figure*}
\centering
\includegraphics[width=0.75\textwidth,trim=0 0 0 0,clip]{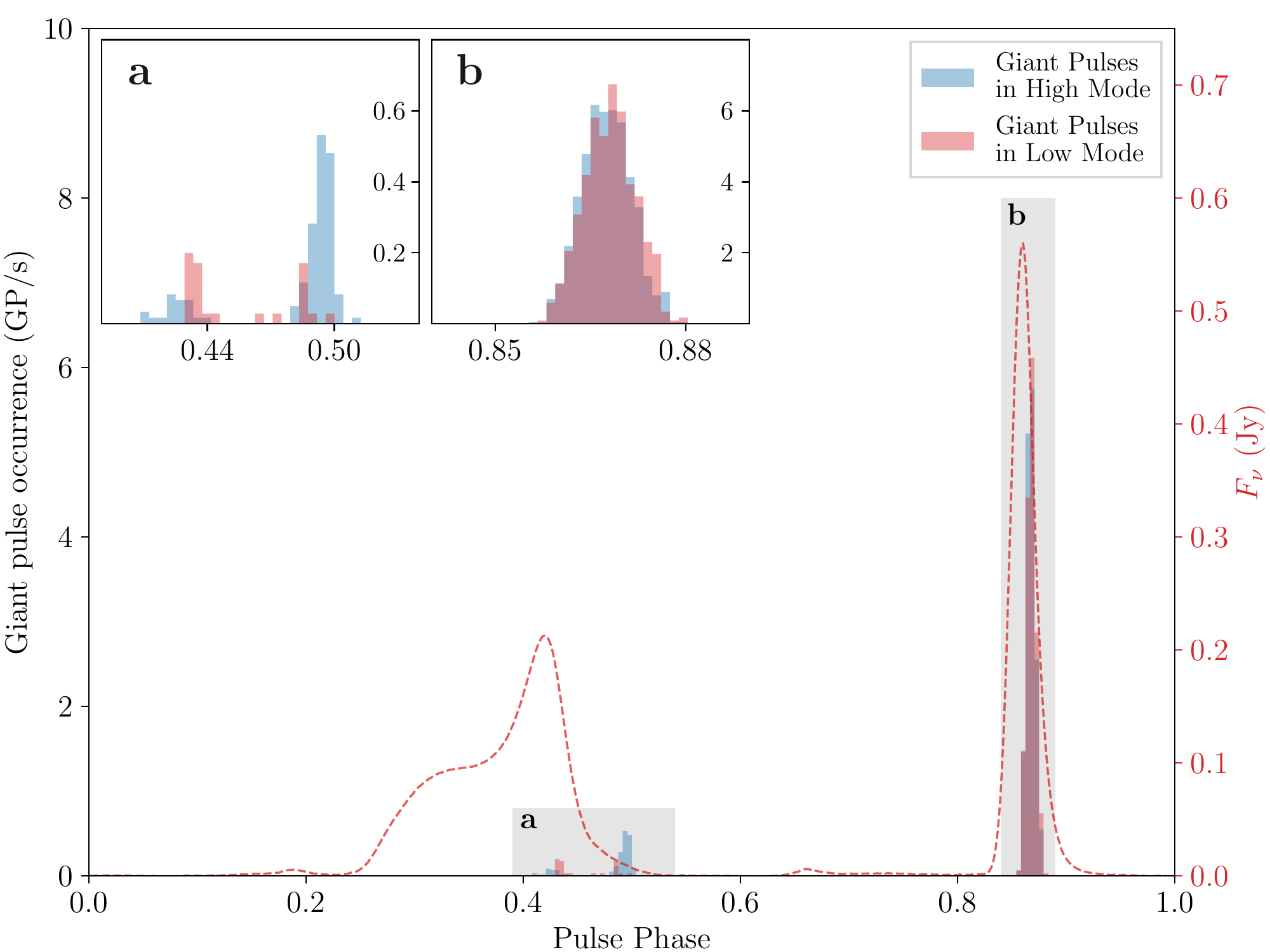}
\caption{Giant pulse occurrence rate as a function of pulse phase and emission mode, with the average pulse profile overlaid (red dashed line, right axis). The rate is per unit time in which the pulsar is at a given phase and in a given mode (i.e., independent of phase binning and mode fraction). The two insets show enlargements for the shaded interpulse (\textbf{a}) and main pulse (\textbf{b}) regions.\label{fig:gp_phases}\\}
\end{figure*}

As described in \cite{Main2017}, we detected hundreds of giant pulses in our data. For our analysis here, we have searched the data again using a more refined detection technique, in which we look for excesses not in a running average of the timestream, but rather in the convolution of the timestream with an exponential decay filter with a scattering timescale, $\tau_s$, that corresponds to the scattering of the regular pulse emission. Assuming that giant pulses are intrinsically extremely short, this should detect them more preferentially over randomly stronger regular pulse emission.

We determine the scattering timescale of the pulsar's emission by fitting the average profile of the main pulse with an exponentially modified Gaussian profile. We measure a scattering timescale of $\tau_s=13.41\pm0.14\,\mu\mathrm{s}$, close to what was inferred from measuring only the brightest giant pulse by \cite{Main2017}.

We select a flux-limited sample by first finding candidates in the convolved timestream at a relatively low signal-to-noise threshold of $10$, and then discarding all for which the integrated flux (over $72\,\mu$s around the giant pulse) is below $100\,\mathrm{Jy}\cdot\mu\mathrm{s}$. For comparison, the integrated flux over the entire main pulse is $25\,\mathrm{Jy}\cdot\mu\mathrm{s}$. This integrated flux can be thought of as a proxy for the giant pulse energy. We detect $1715$ giant pulses using these constraints, with $1575$ near the main pulse and $140$ near the interpulse.

To compare precisely where giant pulses arrive relative to the regular pulse profile, we should take into account that the pulse profile is also convolved by the scattering and thus delayed by the scattering timescale, while times inferred from our exponential fits are not. To compensate for this, we add the scattering timescale to the arrival times before calculating phases.\footnote{This correction was not done in \cite{Main2017}, leading us to conclude, incorrectly, that the giant pulses were coincident with the main pulse.} We determine that the giant pulses in the main pulse are emitted in the trailing half of the pulse components (see Figure~\ref{fig:gp_phases}), in agreement with \cite{Knight2005}, and unlike what is the case for PSR B1937+21, where the giant pulses appear at the extreme trailing ends of the pulse components \citep{Kinkhabwala2000}.

\subsection{Correlation with Mode Changing} \label{ssec:gpcorr}

\begin{figure}
\centering
\includegraphics[width=0.45\textwidth,trim=0 0 0 0,clip]{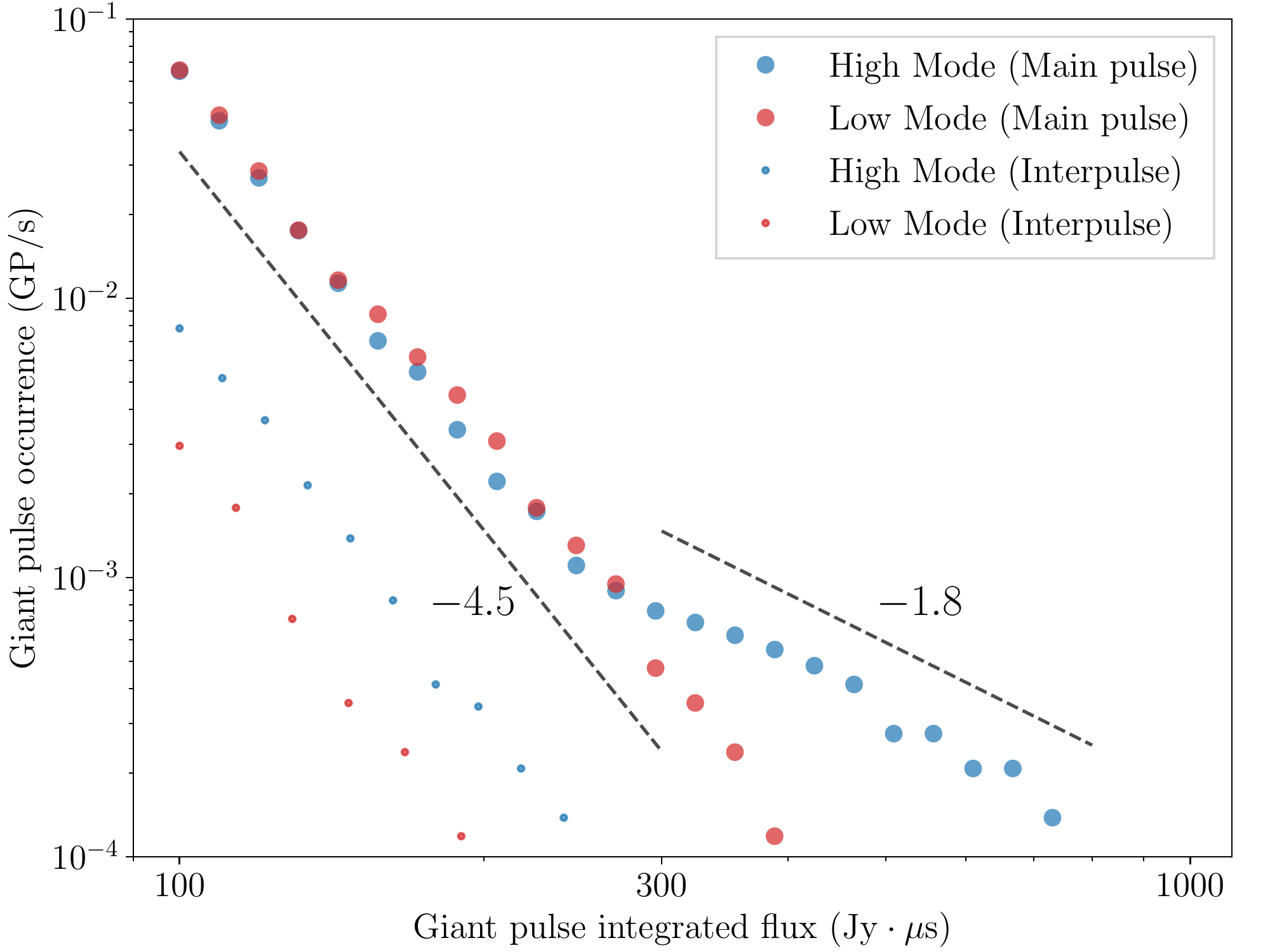}
\caption{Reverse cumulative giant-pulse energy distributions, showing the occurrence rates of giant pulses with integrated flux $E>E_0$ as a function of $E_0$, with giant pulses separated by main pulse and interpulse, and High and Low mode (with the occurrence rate normalized by the fraction of time the pulsar is in each mode). Note the significantly flatter tail for the high-energy main pulse giant pulses in the High Mode. For reference, we show power-law slopes of $-4.5$ and $-1.8$.\label{fig:gp_energy}}
\end{figure}

The distribution of giant pulses with pulse phase is correlated with mode changing. As can be seen in Figure \ref{fig:gp_phases}, the dependence on mode is particularly striking in the interpulse region. In both modes, the distribution is double-peaked, but in the Low mode the pulses arrive in a narrower phase range than is the case for the High mode. Furthermore, the occurrence rate near the trailing edge of the interpulse is much lower in the Low mode.

In the main pulse region, the giant-pulse distributions differ only subtly: that in the Low mode is shifted slightly to later phase compared to that in the High mode by $726\pm344\,\mathrm{ns}$. While not as statistically significant, this is similar to the shift we found for the normal pulse emission suggesting that, for the main pulse, the emission regions responsible for the regular and giant pulses react similarly to the mode changing.

We now turn to the energy distributions of the giant pulses, again considering separately those emitted near the main pulse and interpulse, and in the High and Low mode. As can be seen in Figure~\ref{fig:gp_energy}, at lower energies all four distributions look similar, and seem reasonably well described by a power-law distributions of the form,
\begin{equation}
N_{\mathrm{GP}}(E>E_0)\propto{E_0}^{\alpha},
\end{equation}
with power-law index $\alpha\simeq-4.5$. At higher energies, however, the distribution for the giant pulses emitted near the main pulse in the High mode differs, showing a break to a significantly flatter, $\alpha\simeq-1.8$ power-law distribution.

Comparing these to what is found for the \object{Crab Pulsar} ($\alpha$ from $-1$ to $-3$) and \object{PSR B1937+21} ($\alpha=1.4$; \cite{Bilous2008}), we see that only the high-energy tail of the main-pulse, High-mode giant pulses is comparable. \cite{Bilous2008} found that the energy distributions depends on giant pulse width, with wider pulses having steeper distributions. This could indicate that our high-energy giant-pulse component consists of narrower giant pulses. Unfortunately, the high scattering time prevents us from confirming this directly.

\section{Ramifications} \label{sec:rami}

Our discovery of mode changing in PSR B1957+20 shows that the phenomenon extends to millisecond pulsars. At least in PSR B1957+20, it occurs on a much more rapid timescale, of $1.7\,$s on average, than in regular pulsars, where the timescales range from minutes to weeks. Comparing timescales instead in units of pulse period, the scales are less dissimilar at $\sim\!10^3$ rotations.

The rapid switching may underlie the fact that mode switching had hitherto not been observed in millisecond pulsars: given typical integration times of $10$ s, modes would be averaged out and any left-over residuals might well be attributed to pulse-to-pulse variability or jitter noise. It would be useful to determine whether it exists in other pulsars, however, since it may limit the timing precision achievable if its effects on the pulse arrival time are larger than the measurement error (for data integrated over a typical mode duration). For our Arecibo data of PSR B1957+20, this is the case (although only just): for a profile integrated over the average mode length of $1.7\,$s, the measurement error is about $0.7\,\mu$s, which is smaller than the time offset of $\sim\!1.0\,\mu$s between the two modes. We indeed find that we can improve the timing precision by fitting pulse profiles with a sum of the Low- and High-mode template (with independent amplitudes): for 8\,s integrations, this reduces the RMS of timing residuals from $260$ to $160$\,ns. Of course, at smaller telescopes with noisier data, the effect will be less pronounced. Nevertheless, it may affect the data more surreptitiously. For instance, if the mode fraction were variable over longer timescales, it would lead to systematic errors in pulsar timing experiments that rely on long time integrations yielding stable pulse profiles.

Our observations provide clues to the physical mechanism underlying mode changing. In particular, the fact that the mode changes affect multiple pulse components, both in intensity and polarization, as well as the giant pulse emission, strongly supports the idea that mode changing reflects a global reconfiguration of the pulsar's magnetosphere. If so, the delay of $\sim\!25\,\mathrm{ms}$ between the changes inferred from interpulse and main pulse regions provides a timescale on which mode changing effects propagate across the magnetosphere. It may be useful to compare this with, e.g., the delay between intensity modulations of the main pulse and interpulse of PSR B1055-52 by \cite{Weltevrede2012}. In any case, the delay, as well as the fact that the main pulse is affected much less strongly, also suggest the interpulse is nearer the region responsible for the mode changing.

The detailed pulse profiles and giant-pulse distributions provide additional clues on how the magnetosphere changed. It seems many of the differences between High and Low modes might be understood from changes in geometry. In particular, the weaker interpulse profile and narrower interpulse giant-pulse distribution suggest that in the Low mode the whole interpulse beam moved away from our line of sight. In the main pulse, the slight phase offset of the main pulse, both in regular and giant-pulse emission, again seems suggestive of a change in orientation, albeit only a small one. 

In contrast, the change in energy distribution of the giant pulses associated with the main pulse seems very hard to explain using just geometric arguments, as one has to appeal to a component of giant pulses that is so narrowly aimed and so strongly beamed that even a slight change in orientation of the magnetosphere would make them miss our line of sight. It may be that, instead, a different orientation of the magnetosphere enables a separate type of giant pulse to be emitted.

Follow-up observations at different frequencies might shed further light on the causes of both mode changing and giant pulses. For instance, at higher frequency, scattering is less important and one might be able to tell whether, e.g., the higher-energy giant pulses in High mode differ from the others in their typical duration. Similarly, observations of other millisecond pulsars at lower energy might be fruitful: does a giant-pulse population with a steeper energy distribution appear? And do other millisecond pulsars show mode changing? If so, how does it depend on frequency? Combined, one can hope for further understanding and some movement towards a coherent theory of radio emission from pulsars.

\acknowledgements
We thank Christopher Thompson for helpful discussions, and the referee for useful comments and for pointing out interesting earlier work. We made use of NASA'a Astrophysics Data System and SOSCIP Consortium’s Blue Gene/Q computing platform.

\facility{EVN:Arecibo:327-MHz Gregorian}

\software{
	Astropy \citep{astropy};
	Baseband (\url{http://baseband.readthedocs.io})
}

\bibliographystyle{aasjournal}

\end{document}